\documentclass[reprint,amsmath,amssymb,longbibliography,superscriptaddress,
 aps,prl]{revtex4-2}
\usepackage{ulem}
\usepackage{hyperref}
\usepackage{graphicx}
\usepackage{amsmath}
\usepackage{color}
\usepackage{physics}
\usepackage{pifont}
\usepackage[T1]{fontenc}
\usepackage{multirow}

\def\numbersupplementpages{\the\pdflastximagepages}

\newif\ifarXiv
\arXivfalse

\definecolor{limegreen}{rgb}{0.2, 0.8, 0.2}
\definecolor{orange}{rgb}{1.0, 0.5, 0.0}
\definecolor{blue(ncs)}{rgb}{0.0, 0.53, 0.74}

\definecolor{emerald}{rgb}{0.31, 0.78, 0.47}
\def\red{\color{red}}

\hypersetup{%
     colorlinks=true,
     linkcolor=magenta,
     filecolor=blue,
     citecolor=emerald,      
     urlcolor =blue(ncs),
}

\newcommand{\kv}{{\bf k}}

\begin{document}
\title{Origin of sublattice particle-hole asymmetry in monolayer FeSe superconductors}

\author{Mercè Roig}
\affiliation{Department of Physics, University of Wisconsin-Milwaukee, Milwaukee, Wisconsin 53201, USA}

\author{Kazi Ranjibul Islam}
\affiliation{Department of Physics, University of Wisconsin-Milwaukee, Milwaukee, Wisconsin 53201, USA}

\author{Basu Dev Oli}
\affiliation{Department of Physics and Astronomy, West Virginia University, WV 26506, USA}

\author{Huimin Zhang}
\affiliation{Department of Physics and Astronomy, West Virginia University, WV 26506, USA}

\author{P. M. R. Brydon}
\affiliation{Department of Physics and the MacDiarmid Institute for Advanced Materials and Nanotechnology, University of Otago, P.O. Box 56, Dunedin 9054, New Zealand}

\author{Aline Ramires}
\affiliation{Institute of Solid State Physics, TU Wien, 1040 Wien, Austria}

\author{Yue Yu}
\affiliation{Department of Physics, University of Wisconsin-Milwaukee, Milwaukee, Wisconsin 53201, USA}

\author{Michael Weinert}
\affiliation{Department of Physics, University of Wisconsin-Milwaukee, Milwaukee, Wisconsin 53201, USA}

\author{Lian Li}
\affiliation{Department of Physics and Astronomy, West Virginia University, WV 26506, USA}

\author{Daniel F. Agterberg}
\affiliation{Department of Physics, University of Wisconsin-Milwaukee, Milwaukee, Wisconsin 53201, USA}

%\vskip 1cm

\begin{abstract}
In iron-based superconductors, the two Fe atoms in the unit cell are typically related by crystal symmetries; therefore, we expect no intra-unit cell variations in the superconducting gap. However, recent experiments have challenged this expectation, reporting intra-unit cell variations in the gap with an unusual particle-hole asymmetry. Here, we examine the origin of this asymmetry between the two Fe sublattices in monolayer FeSe grown on SrTiO$_3$. We reveal that, in addition to the substrate-induced broken inversion symmetry, substrate nematic symmetry breaking is key to observing this asymmetry. We further identify two possible mechanisms through which this can occur. The first is through an odd-parity gap function that coexists with an extended $s$-wave function. The second is via a nodeless $d$-wave gap function that develops in the presence of a symmetry-breaking substrate. We argue that the latter mechanism is more physical. To test our theory, we performed scanning tunneling spectroscopy measurements across the nematic domain walls, which exhibit a clear enhancement of the asymmetry between the two Fe sublattices. In addition, we reveal that the observed sublattice particle-hole asymmetry is associated with odd-frequency pairing correlations, providing an experimental realization of this unusual pairing correlation.
\end{abstract}

\date{\today}%

\maketitle

\begin{figure}[t]
\begin{center}
\includegraphics[angle=0,width=\columnwidth]{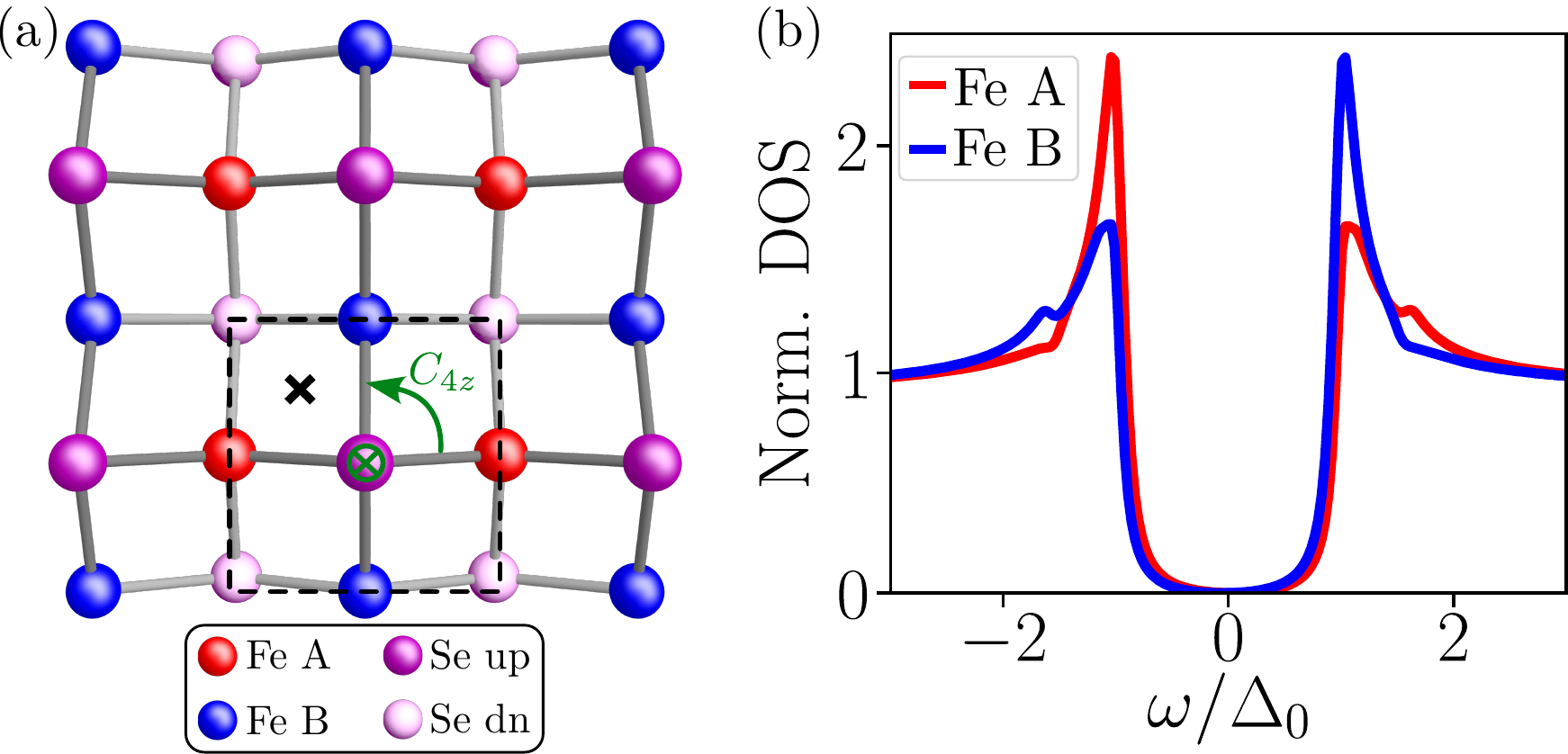}
\caption{(a) Sketch of the monolayer FeSe lattice illustrating the center of inversion (denoted by the X mark) and the four-fold rotation about the $z$ axis ($C_{4z}$) through one Se site relating the A and B Fe sublattices. (b) Schematic representation of the normalized DOS as a function of frequency $\omega$ in units of the superconducting gap $\Delta_0$, displaying a difference between the Fe A and Fe B sublattices.}
\label{fig:schematic_sublatDOS}
\end{center}
\end{figure}

\textit{{\red Introduction.--}} 
Iron-based superconductors have attracted a lot of interest due to the interplay of different electronic orders~\cite{Fernandes2022Jan,Kreisel2020Aug}. The case of monolayer FeSe grown on SrTiO$_3$ (STO) is remarkable, as it exhibits the highest critical temperature among all Fe-based superconductors~\cite{Wang2012Mar,He2013Jul,Zhang2015Jul,Ge2015Mar,Xu2021May}. In the crystallographic unit cell, there exist two symmetry-related Fe atoms, as illustrated in Fig.~\ref{fig:schematic_sublatDOS}(a).
Recently, scanning tunneling microscopy/spectroscopy (STM/STS) has reported measurements in the superconducting state that differ on each Fe site of the unit cell~\cite{Kong2025Apr,Ding2024Jun,Wei2025Feb,Zhang2024Jun}. For monolayer FeSe grown on STO, Ref.~\cite{Ding2024Jun} reported comparable gap magnitudes on the two Fe sites, both exhibiting a nodeless gap, but with the spectra for one Fe site approximately related to the other Fe site by a particle-hole transformation.  This is referred to as a sublattice dichotomy, and is shown in Fig.~\ref{fig:schematic_sublatDOS}(b).

These STM/STS measurements suggest that the two Fe sites are inequivalent due to a symmetry breaking. In particular, since STM is a surface probe, the inversion symmetry relating the two Fe sites is naturally broken. This has led to the suggestion that a substrate-induced odd-parity superconducting order coexisting with an s-wave order underlies this sublattice dichotomy~\cite{Ding2024Jun,Hu2013Jul}. However, this odd-parity superconducting order is an inter-band order, and hence is not stable in the usual weak-coupling limit. 
This motivates the development of a systematic analysis of the origin of the sublattice dichotomy. We carry out such an analysis here. Specifically, by including different patterns of inversion symmetry breaking and different gap functions, we identify the conditions that allow for this sublattice dichotomy. 

Our analysis leads to strong constraints on the nature of the symmetry breaking that allows the sublattice asymmetry to appear. Specifically, we find that it is not sufficient to only break inversion symmetry; a nematic order that also breaks the four-fold symmetry is required. This symmetry breaking is consistent with the observed interfacial structure of FeSe on STO, where domain structures have been observed in which the substrate induces an additional four-fold symmetry breaking consistent with the nematic order found here~\cite{Li2016Mar,Fukaya2018Dec,Peng2020Apr,Zou2025Apr}. Our results point to the importance of understanding this interfacial coupling, which has also been identified as a key factor in the T$_c$ enhancement in this material~\cite{Lee2014Nov,Yang2024Nov,Zou2025Sep}. 

In addition, we show that there are two distinct microscopic mechanisms that allow for this dichotomy to appear. The first is a result of the coexistence of two symmetry-distinct superconducting orders. The second is a result of a single superconducting order, but with the normal electronic state exhibiting the required symmetry breaking. The first mechanism is consistent with that found in Ref.~\cite{Ding2024Jun}, where an odd parity superconducting order coexists with an $s$-wave gap~\cite{Fan2015Nov,Wei2023Aug}. We argue that the second mechanism in which the pairing state corresponds to a nodeless $d$-wave superconducting gap~\cite{Li2016Jun,Agterberg2017Dec,Nakayama2018Dec,Huang2017Mar,Ge2019Apr} provides the most natural explanation. To test our theory, we have measured the tunneling spectra of the two Fe sublattices using atomic-resolution scanning tunneling microscopy/spectroscopy (STM/STS). We find that, as predicted, close to the domain boundary the sublattice dichotomy is large. Such a domain boundary naturally induces the four-fold symmetry breaking predicted to generate the sublattice dichotomy. 
Another conclusion of our analysis is that the observation of this sublattice dichotomy is only possible if odd-frequency pairing correlations develop in the superconducting state~\cite{Black-Schaffer2013Sep,Abrahams1995Jul,Linder2019Dec}, suggesting that this rare pairing state has been observed in monolayer FeSe on STO. 

\textit{{\red Microscopic theory.--}}
Monolayer FeSe has two electron pockets near the M point~\cite{Liu2012Jul}. These pockets are formed from $d_{x^2-y^2}$ and $d_{xz/yz}$ orbitals. Here, we use a symmetry-based $kp$ theory to describe the relevant bands~\cite{Cvetkovic2013Oct,Agterberg2017Dec}. Specifically, we include the four $M$-point bands that are closest to the chemical potential and find an effective theory for the two bands giving rise to the Fermi pockets. The Bloch states for these bands are superpositions of $d_{x^2-y^2}$ and $d_{xz/yz}$ orbitals at the two Fe sublattice sites within the unit cell. An important property of these Bloch states is that they exhibit an orbital-selective behavior: the $d_{x^2-y^2}$ orbitals can be completely Fe-site-localized, while the $d_{xz/yz}$ orbitals cannot~\cite{Roig2025Sep}. As a consequence, the $d_{x^2 - y^2}$ orbitals play the most important role in the observation of any Fe-sublattice asymmetry.

The effective theory for the two bands that cross the chemical potential shares the same symmetry properties as a theory that includes only $d_{x^2-y^2}$ Wannier functions on the Fe sites (details in the Supplementary Material (SM)~\cite{Supplementary}). This is true even though the real Wannier functions are formed from hybridized $d_{x^2-y^2}$ and $d_{xz/yz}$ orbitals. Underlying this theory are the Pauli matrices $\tau_i$ describing the two orbital/sublattice electronic degrees of freedom. In the following, we exploit the symmetry properties of these $\tau_i$ matrices to clarify the origin of the sublattice dichotomy (for comparison with our measured tunneling spectra, we use the full $kp$ theory). A detailed discussion of the origin of these symmetry properties can be found in Refs.~\cite{Cvetkovic2013Oct,Agterberg2017Dec}.  To gain insight into these, we note that the inversion symmetry operator interchanges the two Fe sublattice sites, and here is represented by $\tau_x$. 
This already implies that the operators $\tau_0$ and $\tau_x$ are even parity, while $\tau_y$ and $\tau_z$ are odd-parity. 
Including all symmetry operations, in terms of irreducible representations (IRs) of the point group $D_{4h}$, the symmetry of these operators are given as follows: $\tau_0 \sim A_{1g}$; $\tau_x \sim B_{2g}$, $\tau_y \sim A_{1u}$, and $\tau_z \sim B_{2u}$~\cite{Cvetkovic2013Oct,Agterberg2017Dec}. 
%Notably, $\tau_0$ and $\tau_x$ are of even parity, while $\tau_y$ and $\tau_z$ are of odd parity. 
The normal-state Hamiltonian is given by~\cite{Cvetkovic2013Oct,Agterberg2017Dec,Suh2023Sep}
\begin{equation}
    H_0(\kv) = \varepsilon_{0,\kv} \tau_0 + \gamma_{xy,\kv} \tau_x,
    \label{eq:H_normal_state}
\end{equation}
where the dispersion $\varepsilon_{0,\kv}$ is invariant under $D_{4h}$ (for example, $\varepsilon_{0,\kv} = \frac{\kv^2}{2m}-\mu$) and the term $\gamma_{xy,\kv}$ has $B_{2g}$ symmetry (for example, $\gamma_{xy,\kv}$ $ = \alpha k_x k_y$). The detailed forms are included in the SM~\cite{Supplementary}. Here, we have omitted the spin degree of freedom since we do not include spin-orbit coupling for simplicity.

The presence of the STO substrate breaks inversion symmetry in the normal state. Since the $\tau_z$ operator is odd parity and even under time-reversal symmetry~\cite{Suh2023Sep}, the normal state has an additional contribution due to the interface (the $\tau_y$ operator cannot appear here since it is odd under time-reversal symmetry),
\begin{equation}
    H_I(\kv) =  M_{I,\kv} \tau_z .
    \label{eq:H_interface}
\end{equation}
We analyze in detail the form and momentum dependence of $M_{I,\kv}$ in the following sections.

Now we turn to the superconducting gap structure. The most general superconducting gap function takes the form $\Delta=\sum_{j,k}f_{j,k}(\kv)\tau_j\sigma_k i\sigma_y$~\cite{Suh2023Sep}. Here, we only include $\tau_0$, $\tau_x$, and $\tau_z$ gap functions, since any $\tau_y$ gap anticommutes with the normal state Hamiltonian $H_0+H_I$, and, as a consequence, is purely interband and hence not stable. We further restrict ourselves to spin-singlet gap functions since these are believed to be the most relevant for monolayer FeSe \cite{Huang2017Mar}. Hence, we consider the general gap function
\begin{equation}
    \Delta (\kv) = \Delta_{0,\kv} \tau_0 + \Delta_{x,\kv} \tau_x + \Delta_{z,\kv} \tau_{z}.
    \label{eq:gap_all}
\end{equation}
Importantly, for a given gap symmetry, the momentum dependence of $\Delta_{0,\kv}, \Delta_{x,\kv}$, and $\Delta_{z,\kv}$ will be different due to the different symmetry properties of $\tau_0,\tau_x$, and $\tau_z$. For instance, if $\Delta_{x,\kv}$ is momentum independent, $\Delta_{0,\kv}$ will carry a $k_xk_y$ dependence to belong to the same representation. If instead $\Delta_{x,\kv}\sim k_x k_y$, then $\Delta_{0,\kv}$ will be momentum independent. The odd-parity gap $\Delta_{z,\kv} \tau_z$ has a momentum-even $\Delta_{z,\kv}$ since $\tau_z$ is parity odd. We initially consider the general form for the superconducting gap in Eq.~\eqref{eq:gap_all} to examine the sublattice asymmetry, and we carry out a detailed analysis of the gap symmetries and the corresponding momentum dependence later.

\textit{{\red Origin of the sublattice asymmetry.--}}
Considering the previous normal-state Hamiltonian and superconducting gaps, the Bogoliubov-de Gennes (BdG) Hamiltonian can be written as
\begin{equation}
    \begin{aligned}
    H_{\rm BdG} (\kv)= & \, \rho_z (\varepsilon_{0,\kv} \tau_0 + \gamma_{xy,\kv} \tau_x + M_{I,\kv} \tau_z) \\
    &+ \rho_x(\Delta_{0,\kv} \tau_0 + \Delta_{x,\kv} \tau_x + \Delta_{z,\kv} \tau_z),
    \end{aligned}
    \label{eq:BdG_sublattice}
\end{equation}
with $\rho_i$ the Pauli matrices denoting particle-hole space. In the band basis, superconducting fitness~\cite{Ramires2016Sep,Ramires2018Jul} implies a finite interband pairing will arise if one or both of the following terms is nonzero: $M_{I,\kv} \Delta_{x,\kv}$ or $\gamma_{xy,\kv} \Delta_{z,\kv}$, see the SM~\cite{Supplementary}. This interband pairing plays a key role in generating the sublattice dichotomy.
Using the previous BdG Hamiltonian, we can obtain an analytic result for the Green's function in the sublattice basis by calculating
\begin{equation}
    \begin{pmatrix}
        \mathcal{G}^e & \mathcal{F} \\
        \mathcal{F}^\dagger & \mathcal{G}^h
    \end{pmatrix} 
    = [i\omega_n \rho_0\tau_0 - {H}_{\rm BdG}(\kv)]^{-1},
    \label{eq:Greens_function_general}
\end{equation}
including the electron ($\mathcal{G}^e$) and the hole ($\mathcal{G}^h$) part, and the anomalous Green's function ($\mathcal{F}$). The A and B sublattice-resolved density of states (DOS) then correspond to~\cite{Kong2025Apr,Papaj2025Jun}
\begin{equation}
    \mathcal{N}_{A/B}(\omega)\! =\! -\frac{1}{\pi}\Im\! \sum_\kv\! \frac{1}{2} (\tau_0 \pm \tau_z)\mathcal{G}^e(i\omega_n\rightarrow \omega+i\eta,\kv). \label{eq:DOS}
\end{equation}
Here, the $+$ ($-$) denotes the projection to the A (B) sublattice.

To determine the relevant parameters yielding a sublattice dichotomy, we obtain an analytic expression for the Green's function particle-hole asymmetry between the two sublattices (see the SM~\cite{Supplementary}). There are two contributions to the particle-hole asymmetry: one from the normal state ($\mathcal{G}_{\rm asym}^{N}$) and one from the superconducting state ($\mathcal{G}_{\rm asym}^{SC}$)~\cite{Supplementary}. Here, we focus on the asymmetry arising from the superconducting state because it is numerically larger, 
\begin{equation}
    \mathcal{G}_{\rm asym}^{\rm SC} \!(\kv,\omega_n) {=}\!\!  \sum_{\nu=\pm} \!f(\omega_n) (M_{I,\kv} \Delta_{x,\kv} {-} \gamma_{xy,\kv} \Delta_{z,\kv}) \tilde{\Delta}_{x,\kv},
    \label{eq:sublat_difference} 
\end{equation}
where $\tilde{\Delta}_{x,\kv} = \Delta_{x,\kv}  + \frac{ \nu \Delta_{0,\kv}\gamma_{xy,\kv}}{\sqrt{\gamma_{xy,\kv}^2+M_{I,\kv}^2}}$ carries the same symmetry as the gap $\Delta_{x,\kv}$. In addition, $f(\omega_n) = \frac{1}{(\omega_n^2+E_+^2)(\omega_n^2+E_{-}^2)} \frac{\omega_n^2}{\omega_n^2 +\xi_\nu^2} $, $\nu$ indexes the two normal-state bands, $\xi_{\pm}= \varepsilon_{0,\kv} \pm \sqrt{\gamma_{xy,\kv}^2+M_{I,\kv}^2}$, and $E_{\pm}$ is the quasi-particle spectrum of the BdG Hamiltonian in Eq.~\eqref{eq:BdG_sublattice}~\cite{Supplementary}.

To obtain a finite sublattice asymmetry in the DOS, the sum over momenta in Eq.~\eqref{eq:DOS} must not vanish. Importantly, this reveals two possible scenarios. On the one hand, if an odd parity gap $\Delta_{z,\kv}$ coexists with an even parity gap $\tilde{\Delta}_{x,\kv}$, the DOS for the A and B sublattices is different provided that the combined symmetry $\gamma_{xy,\kv} \Delta_{z,\kv}\tilde{\Delta}_{x,\kv}$ is $A_{1g}$, so that the momentum sum in the DOS does not vanish. On the other hand, a sublattice difference can also emerge from the substrate-induced symmetry breaking term if the momentum sum of $M_{I,\kv}\Delta_{x,\kv}\tilde{\Delta}_{x,\kv}$ does not vanish, which implies that $M_{I,\kv}$ should be momentum independent.
In the following sections, we analyze the symmetries of the interface symmetry breaking and the superconducting gaps.

\begin{figure}[t]
\begin{center}
\includegraphics[angle=0,width=0.8\columnwidth]{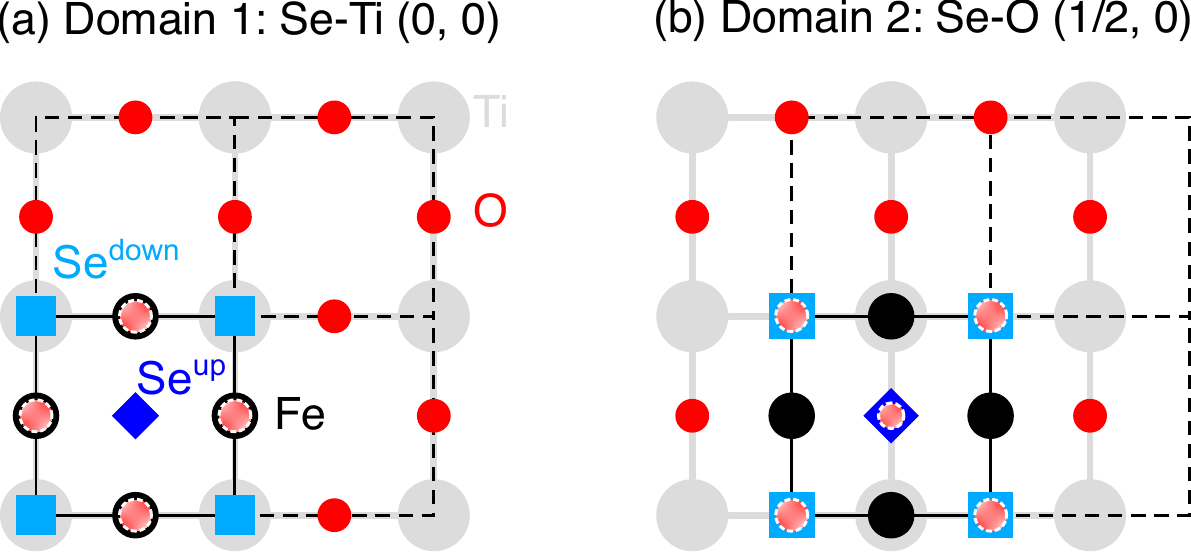}
\caption{Two domains of monolayer FeSe grown on STO, inducing different substrate symmetry breakings. The O atoms that are below Fe or Se atoms are illustrated by a dashed circle with a light red interior. (a)~The mirror symmetry $m_z$ relating the two Fe sublattices is broken. (b)~Only one Fe sublattice is on the Ti atoms and, consequently, both the mirror and four-fold symmetry relating the two Fe sites are broken.}
\label{fig:domains_STO}
\end{center}
\end{figure}

\textit{{\red Interface symmetry breaking.--}}
In the presence of the STO interface, depending on the position of the substrate atoms with respect to the monolayer FeSe lattice, different symmetry breakings can be induced. As illustrated in Fig.~\ref{fig:domains_STO}, we focus on two different domains observed experimentally~\cite{Peng2020Apr,Zou2025Apr}. In the first case, shown in Fig.~\ref{fig:domains_STO}(a), the Se atoms in the upper layer lie in the hollow of the TiO$_x$ layer, while the Se atoms in the bottom layer are on Ti atoms. In this domain, the mirror symmetry $m_z$ relating the two Fe sublattices is broken. Therefore, an $A_{2u}$ symmetry-breaking term transforming like $k_z$ is induced in the normal state Hamiltonian. In the second type of domain, illustrated in Fig.~\ref{fig:domains_STO}(b), the Se atoms are on O atoms and only one Fe sublattice (see Fig.~\ref{fig:schematic_sublatDOS}(a)) lies on the Ti atoms. As a consequence, both the mirror and the four-fold symmetry relating the two Fe sublattices are broken, rendering the two sublattices inequivalent and giving rise to a different DOS. 
In particular, the four-fold symmetry breaking generates a nematic $\epsilon_{x^2-y^2}$ order, introducing an additional normal-state symmetry breaking transforming like $B_{2u}\sim k_z(k_x^2-k_y^2)$.

In our effective $kp$ model, $\tau_z$ transforms like the $B_{2u}$ IR, introducing a momentum-independent $B_{2u}$ symmetry breaking term. However, the presence of the $A_{2u}$ symmetry breaking induces a term that is momentum dependent, transforming like $(k_x^2-k_y^2)\tau_z$ (since $B_{1g} \otimes B_{2u} = A_{2u}$). Therefore,including both $A_{2u}$ and $B_{2u}$ symmetry breaking, the normal-state symmetry breaking entering in Eq.~\eqref{eq:H_interface} corresponds to
\begin{equation}
    M_{I,\kv} =  M_{A_{2u}} (k_x^2-k_y^2)+M_{B_{2u}} ,
    \label{eq:M_sym}
\end{equation}
where $M_{A_{2u}}$ and $M_{B_{2u}}$ are the constants determining the strength of the two symmetry breaking terms. 
Equation~\eqref{eq:DOS} shows that there is no sublattice asymmetry for an $A_{2u}$ symmetry breaking, since the momentum sum in the DOS vanishes for odd powers of $M_{A_{2u}} (k_x^2-k_y^2)$. A similar argument reveals that there is no sublattice asymmetry from $B_{1u}$, $A_{1u}$, or $E_u$ interface symmetry breaking. This leads to one of our key conclusions: only a $B_{2u}$ interface symmetry breaking can allow a sublattice asymmetry. 

\begin{table}[t]
\caption{Irreducible representation of the total superconducting order parameter in Eq.~\eqref{eq:gap_all}. The nodeless $d$-wave ($B_{2g}$ gap) corresponds to $\Delta_{x,\kv} \sim c $, while the extended $s$-wave ($A_{1g}$ gap) corresponds to $\Delta_{x,\kv} \sim k_x k_y$.}
\label{tab:gaps}
\begin{tabular}{ccccccc} \hline \hline
 & $c,k_x^2+k_y^2$ & $k_x^2-k_y^2$ & $k_x k_y$  \\ \hline
$\tau_0$ & $A_{1g}$ & $B_{1g}$ & $B_{2g}$  \\
$\tau_x$ & $B_{2g}$ & $A_{2g}$ & $A_{1g}$ \\
$\tau_z$ & $B_{2u}$ & $A_{2u}$ & $A_{1u}$  \\ \hline \hline
\end{tabular}
\end{table}

\textit{{\red Superconducting gaps.--}}
To derive the sublattice particle-hole asymmetry in Eq.~\eqref{eq:sublat_difference}, we have considered the superconducting order parameter as in Eq.~\eqref{eq:gap_all}. We now examine the symmetries of the different spin-singlet gaps. Since $\tau_0$ and $\tau_x$ are parity even, the symmetry of the total superconducting gap (including $\Delta_{0,\kv}$ and $\Delta_{x,\kv}$, respectively) is also even under inversion, while the total symmetry of the gap $\Delta_{z,\kv} \tau_z$ is of odd parity. In Table~\ref{tab:gaps}, we include the IR of the total superconducting order parameters in Eq.~\eqref{eq:gap_all}.
Note that there are two terms transforming like a $B_{2g}$ order parameter: the nodeless $d$-wave gap $c\, \tau_x$ and an on-site gap $k_x k_y \tau_0$. Similarly, the $A_{1g}$ order parameter also includes the extended $s$-wave gap $k_x k_y \tau_x$ and an on-site term $c\, \tau_0$.

\begin{figure}[t]
\begin{center}
\includegraphics[angle=0,width=\columnwidth]{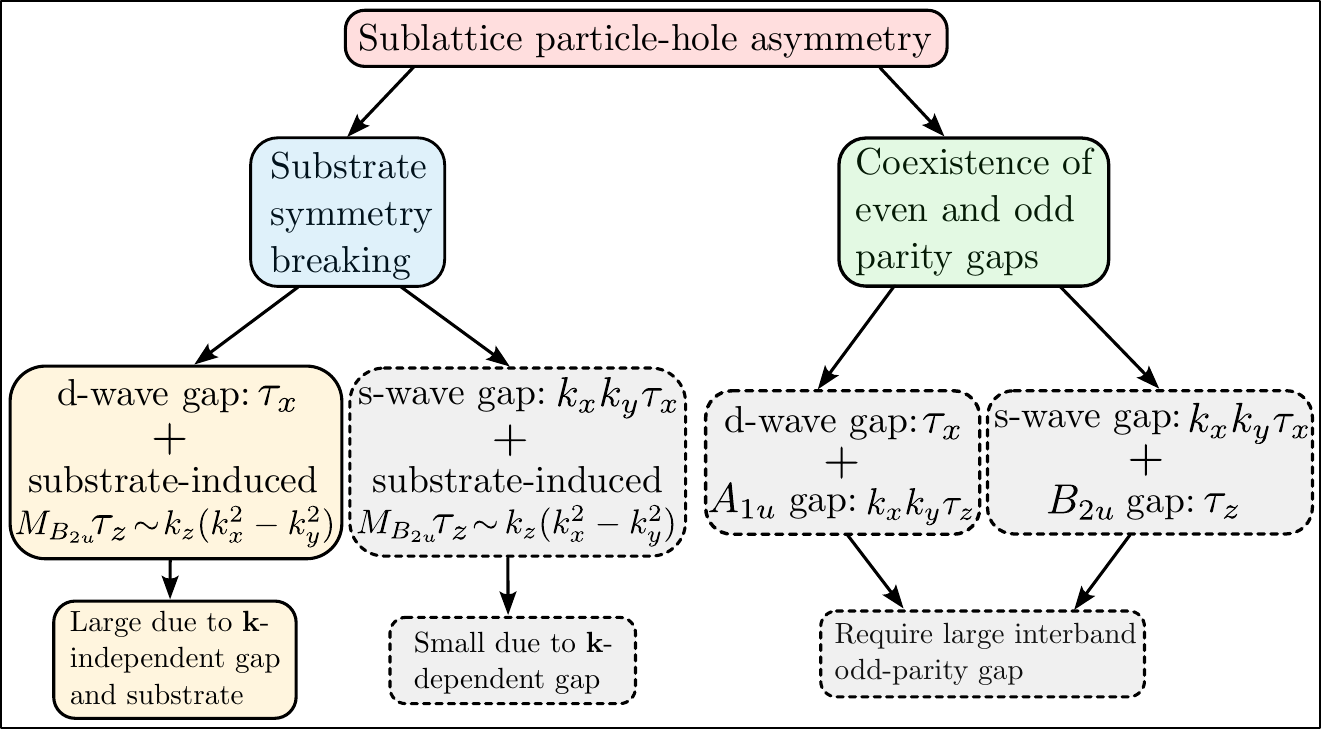}
\caption{Map showing the two scenarios that give rise to a sublattice particle-hole asymmetry for the nodeless $d$-wave and extended $s$-wave gap.}
\label{fig:schematic_asymmetry}
\end{center}
\end{figure}

In Fig.~\ref{fig:schematic_asymmetry}, we show the two different scenarios giving rise to a sublattice asymmetry.
Importantly, the second term in Eq.~\eqref{eq:sublat_difference} shows that, in the absence of $M_{I,\kv}$, a sublattice asymmetry can also emerge from the coupling of an even parity and an odd parity gap. Specifically, focusing on the extended $s$-wave $A_{1g}$ gap $ k_x k_y \tau_x$, and since $\gamma_{xy,\kv}\sim k_x k_y$, there is a sublattice asymmetry only in the presence of a momentum-independent $B_{2u}$ gap $\Delta_{z,\kv} \tau_z \sim c\, \tau_z$, as proposed in Ref.~\cite{Ding2024Jun}. 
Considering now the nodeless $d$-wave $B_{2g}$ gap $c\,\tau_x$, a sublattice asymmetry is also generated if it coexists with an odd parity $A_{1u}$ gap $\Delta_{z,\kv} \tau_z \sim k_x k_y \tau_z$.

However, in the absence of a substrate-induced symmetry breaking $M_{I,\kv}$, the odd-parity gaps $B_{2u}$ and $A_{1u}$ are purely interband, and therefore they are unlikely to be intrinsically stabilized and would require two separate superconducting transitions to appear. In contrast, the difference between the two sublattices naturally arises from the substrate-induced symmetry breaking, see Fig.~\ref{fig:schematic_asymmetry}. As seen from Eqs.~\eqref{eq:DOS}-\eqref{eq:sublat_difference}, the term $M_{B_{2u}} \Delta_{x,\kv} $ in principle gives rise to a sublattice asymmetry for both the nodeless $d$-wave and the extended $s$-wave gap. However, due to the $k_x k_y$ momentum dependence in the extended $s$-wave gap, the effect is remarkably small. In addition, the pure extended s-wave gap exhibits a V-shape DOS, which is inconsistent with the experimental data. 

\begin{figure*}[t]
\begin{center}
\includegraphics[angle=0,width=\linewidth]{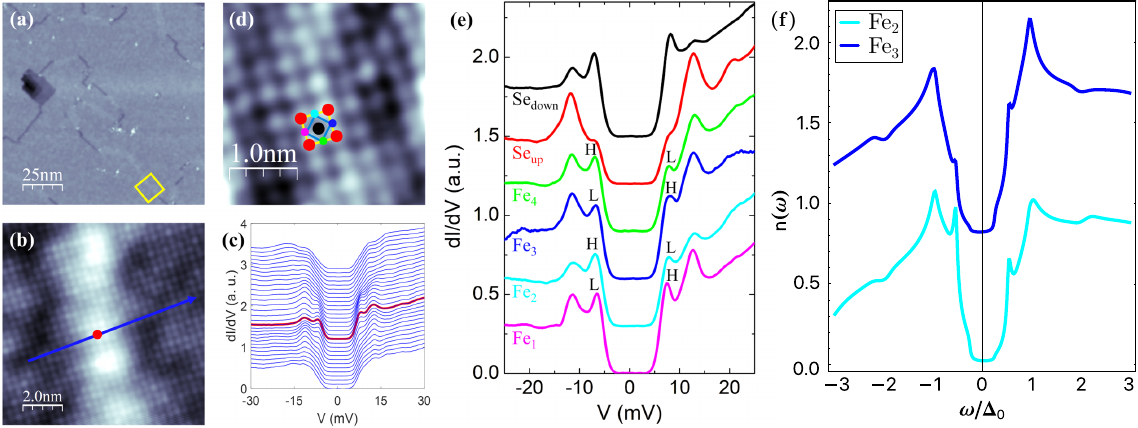}
\caption{STM/STS of the single-layer FeSe/STO(100) and sublattice resolved DOS from the effective model. (a) Large-scale STM image showing two types of domain boundaries orientated along the [10] and [11] directions  (V$_{\rm bias}$ = 3V, I = 10 pA). (b) Atomic resolution image acquired at the boxed area in (a), V$_{\rm bias}$ = 75 mV, I = 100 pA. (c) A series of dI/dV tunneling spectra acquired along the blue arrow in (b) across a [10] domain boundary, V$_{\rm bias}$ = 30 mV, I = 500 pA, V$_{\rm mod}$ = 0.6 mV, tunneling spectra are vertically shifted for clarity. (d) Atomic resolution STM image of the single-layer FeSe/STO(100) (V$_{\rm bias}$ = 30 mV, I = 100 pA), including the ball-stick model showing the top Se lattice (red), four-iron sites and bottom Se (black). (e) dI/dV tunneling spectra acquired at sites indicated in (d), V$_{\rm bias}$ = 30 mV, I = 500 pA, V$_{\rm mod}$ = 0.6 mV, tunneling spectra are vertically shifted for clarity. H and L denote the high and low peaks, respectively. (f) Sublattice-resolved DOS considering the full model including $d_{x^2-y^2}$ and $d_{xz,yz}$ orbitals and the hybridization between them, see the SM~\cite{Supplementary}.}
\label{fig:Fig_tunneling}
\end{center}
\end{figure*}

\textit{{\red Tunneling spectra.--}} 
By examining the interface symmetry breaking and the superconducting gaps entering in Eq.~\eqref{eq:sublat_difference}, it is evident that, in addition to inversion symmetry breaking, a four-fold rotational symmetry breaking $k_x^2-k_y^2$ is necessary to induce a sublattice particle-hole asymmetry. To explicitly demonstrate that, we have performed STM/STS on monolayer FeSe on STO to measure the sublattice asymmetry close to a [10] domain boundary, which effectively introduces $k_x^2-k_y^2$ nematicity, naturally breaking the four-fold symmetry and introducing an asymmetry between the two sublattices. This scenario is similar to the second domain presented in Fig.~\ref{fig:domains_STO}(b). 

As seen from Fig.~\ref{fig:Fig_tunneling}, we identified the two type of domain boundaries, characterized respectively by high and low contrast, and obtained a series of dI/dV tunneling spectra across one low contrast boundary, see Fig.~\ref{fig:Fig_tunneling}(b). We determined the atoms position from the atomic resolution STM image in Fig.~\ref{fig:Fig_tunneling}(d), and obtained the dI/dV spectra at the different sites, see Fig.~\ref{fig:Fig_tunneling}(e). Importantly, close to the domain boundary, the difference between the two sublattices becomes notably larger. Fig.~\ref{fig:Fig_tunneling}(e) also demonstrates the particle-hole asymmetry between the two Fe sites. Specifically, the green and light blue sites in Fig.~\ref{fig:Fig_tunneling}(d) show a higher (lower) peak for negative (positive) voltages, while the opposite occurs for the purple and dark blue sites.
In order to compare with the tunneling spectra measurements, we have considered the full model for the effective theory near the M point including $d_{x^2-y^2}$ and $d_{xz/yz}$ orbitals. As seen from Fig.~\ref{fig:Fig_tunneling}(f), the full model reproduces the main experimental features, including the higher and lower peaks particle-hole asymmetry for the two Fe sublattices.

\textit{{\red Odd frequency pairing.--}} The origin of the particle-hole asymmetry can be ascribed to the presence of odd-frequency pairing correlations. This unusual type of pairing correlations has been proposed in the context of multiband superconductors~\cite{Black-Schaffer2013Sep} and was later associated with the concept of superconducting fitness and finite inter-band pairing~\cite{Ramires2016Sep,Ramires2018Jul}.
We can obtain a general expression for the even and odd-frequency electron Green's function~\cite{Supplementary}. Focusing on the odd-frequency terms, 
\begin{equation}
    \mathcal{G}_{\rm e,asym} = \mathcal{G}^{(0)}_{\rm e,asym} + \mathcal{F}_{\rm sym} \Delta^\dagger \mathcal{G}^{(0)}_{\rm e,asym} + \mathcal{F}_{\rm asym} \Delta^\dagger \mathcal{G}^{(0)}_{\rm e,sym},
    \label{eq:asym_G}
\end{equation}
where $ \mathcal{G}^{(0)}_{\rm e,sym}$/$\mathcal{G}^{(0)}_{\rm e,asym}$ encodes the normal-state particle-hole symmetry/asymmetry in the DOS, $\frac{1}{2}[\mathcal{N}(\omega) \pm \mathcal{N}(-\omega)]$, and $\mathcal{F}_{\rm sym/asym} = \frac{1}{2}[\mathcal{F}(\kv,\omega)\pm \mathcal{F}(\kv,-\omega)]$ are the even and odd frequency components of the anomalous Green's function in Eq.~\eqref{eq:Greens_function_general}. The analytic expressions for these terms are given in the SM~\cite{Supplementary}. Importantly, Eq.~\eqref{eq:asym_G} reveals that we can connect the odd-frequency pairing induced by the superconducting gap to the asymmetry in the sublattice-resolved DOS. Since the asymmetry introduced by the first two terms in Eq.~\eqref{eq:asym_G} arise from the normal-state Green's function, the odd frequency part from the superconducting state originates from the odd frequency anomalous Green's function in the last term. 

By examining the model for FeSe introduced in Eq.~\eqref{eq:BdG_sublattice}, we can obtain an expression for the odd frequency anomalous Green's function,
\begin{equation}
    \mathcal{F}_{\rm asym} = \frac{2\big[- M_{I,\kv} \Delta_{x,\kv} + \gamma_{xy,\kv} \Delta_{z,\kv} \big] \, \omega_{n} \tau_y}{(\omega_n^2+E_+^2)(\omega_n^2+E_-^2)},
\end{equation}
where $E_{\pm}$ the quasi-particle spectrum of the BdG Hamiltonian in Eq.~\eqref{eq:BdG_sublattice}~\cite{Supplementary}. Notably, obtaining a finite $\mathcal{F}_{\rm asym}$ requires the same condition as interband pairing or a finite superconducting fitness~\cite{Ramires2016Sep,Ramires2018Jul}.
Therefore, the previous expressions reveal that in the presence of an interface-symmetry breaking, the sublattice particle-hole asymmetry for the nodeless $d$-wave gap originates from the substrate-induced odd frequency pairing. Similarly, in the absence of $M_{I,\kv}$, and odd parity gap $\Delta_{z,\kv}$ also induces odd frequency pairing. 

\textit{{\red Conclusions.--}} 
We have identified the conditions that allow for a sublattice particle-hole asymmetry in monolayer FeSe. By examining analytic expressions for the Green's function, we have identified two possible scenarios yielding an asymmetry between the two Fe sublattices. In the first case, we show that a substrate-induced inversion symmetry breaking that also breaks a four-fold symmetry is required (generating a non-zero nematic $\epsilon_{x^2-y^2}$-like order). Within this scenario, we have revealed that the sublattice asymmetry is larger for the nodeless $d$-wave case, while significantly smaller for the extended $s$-wave due to the momentum dependence of the gap. In the second scenario, the coexistence of the $d$-wave and $s$-wave with odd parity gaps may also induce a sublattice asymmetry, even though in the absence of a substrate-induced symmetry breaking these gaps are purely interband and unlikely to be intrinsically stabilized. 

Our tunneling spectra measurements provide further evidence for the necessary four-fold symmetry breaking by demonstrating that, close to a [10] domain boundary, this naturally introduces $k_x^2-k_y^2$ nematicity and a large sublattice particle-hole asymmetry is observed. Finally, we have revealed that the sublattice asymmetry originates from a substrate-induced odd-frequency pairing, providing experimental evidence for this elusive type of pairing correlations.

\textit{{\red Acknowledgements.--}}
Work at UWM and WVU was supported by National Science Foundation Grants No.\ DMREF 2323857 and No.\ DMREF 2323858, and by the U.S. Department of Energy, Office of Basic Energy Sciences, Division of Materials Sciences and Engineering, under Award No. DE-SC0017632.
D.F.A, A.R., M.R., K.R.I., and Y.Y. acknowledge support from the Simons Foundation grant SFI-MPS-NFS-00006741-02.
\bibliography{FeSe_LDOS}

\end{document}